\documentclass[twocolumn,prl,unsortedaddress,superscriptaddress]{revtex4}
\usepackage{amsmath,color}
\usepackage[pdftex]{graphicx}

\begin{document}
\title{Acoustic metamaterial exhibiting four different sign combinations of density and modulus}

\vspace{\baselineskip}

\author{Yong Mun Seo}
\affiliation{Department of Physics, Myongji University, Yongin
449-728, Korea}
\author{Jong Jin Park }
\affiliation{Institute of Physics and Applied Physics, Yonsei
University, Seoul 120-749, Korea}
\author{Seung Hwan Lee}
\affiliation{Institute of Physics and Applied Physics, Yonsei
University, Seoul 120-749, Korea}
\author{Choon Mahn Park}
\affiliation{AEE Center, Anyang University, Anyang 430-714, Korea}
\author{Chul Koo Kim}
\affiliation{Institute of Physics and Applied Physics, Yonsei
University, Seoul 120-749, Korea}
\author{Sam Hyeon Lee\footnote{e-mail:
samlee@yonsei.ac.kr}} \affiliation{Institute of Physics and Applied
Physics, Yonsei University, Seoul 120-749, Korea}
\date\today

\vspace{\baselineskip}

\begin{abstract}
\textbf{We fabricated a double negative acoustic metamaterial which
consisted of Helmholtz resonators and membranes. Experimental data
on the transmission and dispersion relation are presented. The
system exhibits three frequencies where the acoustic state makes
sharp transitions from density negative ($\rho$ -NG) to double
negative (DNG), modulus negative ($B$-NG), and double positive (DPS)
in sequence with the frequency. We observed a wide range of negative
refractive index from -0.06 to -3.7 relative to air, which will
allow for new acoustic transformation techniques.}
\end{abstract}

\maketitle

For electromagnetic media, the double negative quadrant of the
$\epsilon$-$\mu$ diagram suggested by Veselago became reality when a
composite structure consisting of split-ring resonators and
conducting wires was proposed by Pendry \emph{et al.} and fabricated
by Smith \emph{et al.}~\cite{1,2,3,4}. This metamaterial became a
standard model for a variety of double negative
media~\cite{5,6,7,8}. Constitutive parameters for acoustic media are
density ($\rho$) and modulus ($B$). In previous research, an
acoustic metamaterial with a negative modulus was fabricated using
an array of Helmholtz resonators~\cite{9}. A negative modulus has
also been generated with an array of side holes~\cite{10}. A
negative density was realized using double resonant units in
fluids~\cite{11,12,13}. Recently, a negative density was also
generated with thin tight membranes~\cite{14}. Simultaneously
negative density and modulus was first realized in a composite
structure consisting of membranes and side holes. This structure
exhibited wide double negative spectral range, but was not capable
of generating large negative refractive index. In this Letter, we
report fabrication of a new acoustic DNG structure which exhibited
all the four different sign combinations of density and modulus, and
a very wide range of negative refractive index. The structure, shown
in Fig. 1, consisted of an array of interspaced Helmholtz resonators
and membranes.

The main tube was made of plastic, and the method outlined in Refs.
14 and 15 was used to fabricate the membranes. Commercial 100 mL
glass bottles were used as the Helmholtz resonators. The bottles
were attached to the tube by plastic adaptors. To create leak-tight
seals, the adapters were machined to fit both the bottle neck and
the side holes made on the tube wall. The size of the neck of the
Helmholtz resonator was determined by the diameter (10 mm) and
length (15 mm) of the inner hole of the adaptor. The unit cell
length of the metamaterial was 70 mm and the inner diameter of the
tube was 32.5 mm.

Without the membranes, the structure in Fig. 1 consists solely of an
array of Helmholtz resonators, which exhibits an effective modulus
$B_{eff}(\omega)$ expressed as~\cite{9},

\begin{figure}
\begin{center}
\includegraphics*[width=1.0\columnwidth]{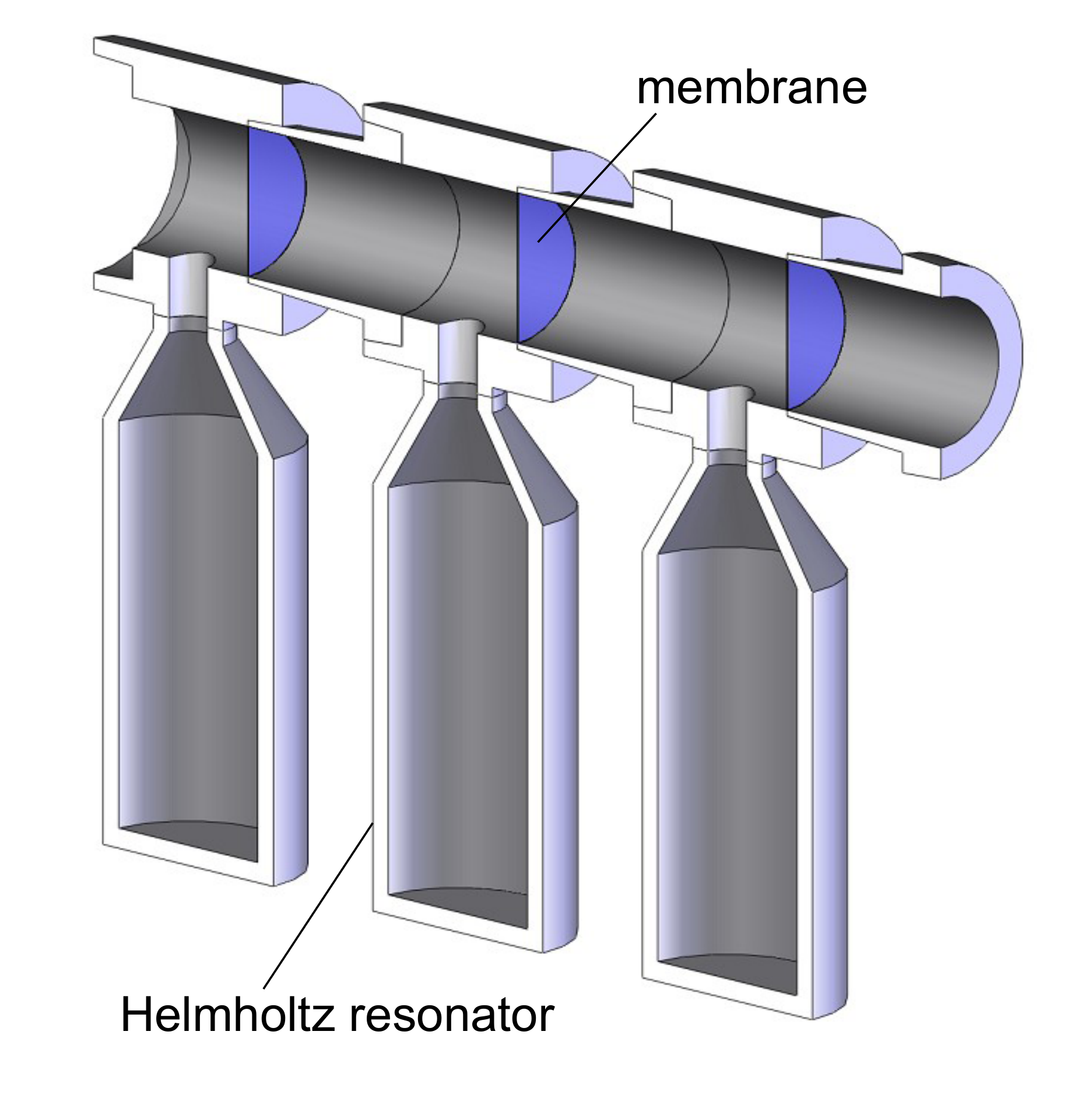}
\end{center}
\caption{Composite structure consisting of an array of interspaced
membranes and Helmholtz resonators. This structure exhibited both a
negative density and a negative modulus.} \label{fig:structure}
\end{figure}
\begin{equation*} \label{eq:rho1}
B_{eff}^{-1}(\omega)=B_0^{-1}\left(1-\frac{F\omega_0^2}{\omega^2-\omega_0^2+i\Gamma\omega}\right),
\end{equation*}
where $B_0$, $F$, $\omega_0$, and $\Gamma$ are the modulus of the
background fluid, the geometric factor, the resonance frequency of
the Helmholtz resonator, and the dissipation factor, respectively.
In the system detailed in Ref. 9, the dissipation was very large
$(\omega_0<\Gamma)$ and thus, a negative group velocity was
observed. The much larger dimensions and better quality factor of
the Helmholtz resonators ($Q\approx 20$) in our structure make the
overall dissipation quite small $(\omega_0\gg\Gamma)$.The
experimental results presented in this Letter can be aptly explained
by neglecting $\Gamma$,

\begin{equation} \label{eq:kai}
B_{eff}^{-1} = B_0^{-1}\left(1-
\frac{F\omega_0^2}{\omega^2-\omega_0^2}\right).
\end{equation}
This equation shows that the effective modulus is negative in the
frequency range $\omega_0<\omega<\omega_0\sqrt{1+F}$.

Without Helmholtz resonators, our system is identical to that of a
$\rho$ -NG structure consisting of an array of membranes~\cite{14}.
In previous research, the $\rho$ -NG structure was found to convey a
collective acoustic oscillation that was similar to a plasma
oscillation. In addition, the structure exhibited an effective
density $\rho_{eff}$ given by,

\begin{equation} \label{eq:rho}
\rho_{eff}(\omega) =
\rho'\left(1-\frac{\omega^2_c}{\omega^2}\right),
\end{equation}
where $\rho'$, the density of membrane-loaded air, is a constant
that is larger than the density of air $\rho_0$. It should be noted
that the frequency dependence of Eq. (2) has the same form as that
of the electrical permittivity of metals,
$\epsilon(\omega)=\epsilon_{0}(1-\omega_p^2/\omega^2)$, where
$\omega_p$ is the plasma frequency. The effective density of the
acoustic metamaterial is negative below the cutoff frequency
$\omega_c$.

\begin{figure}
\begin{center}
\includegraphics*[width=1.0\columnwidth]{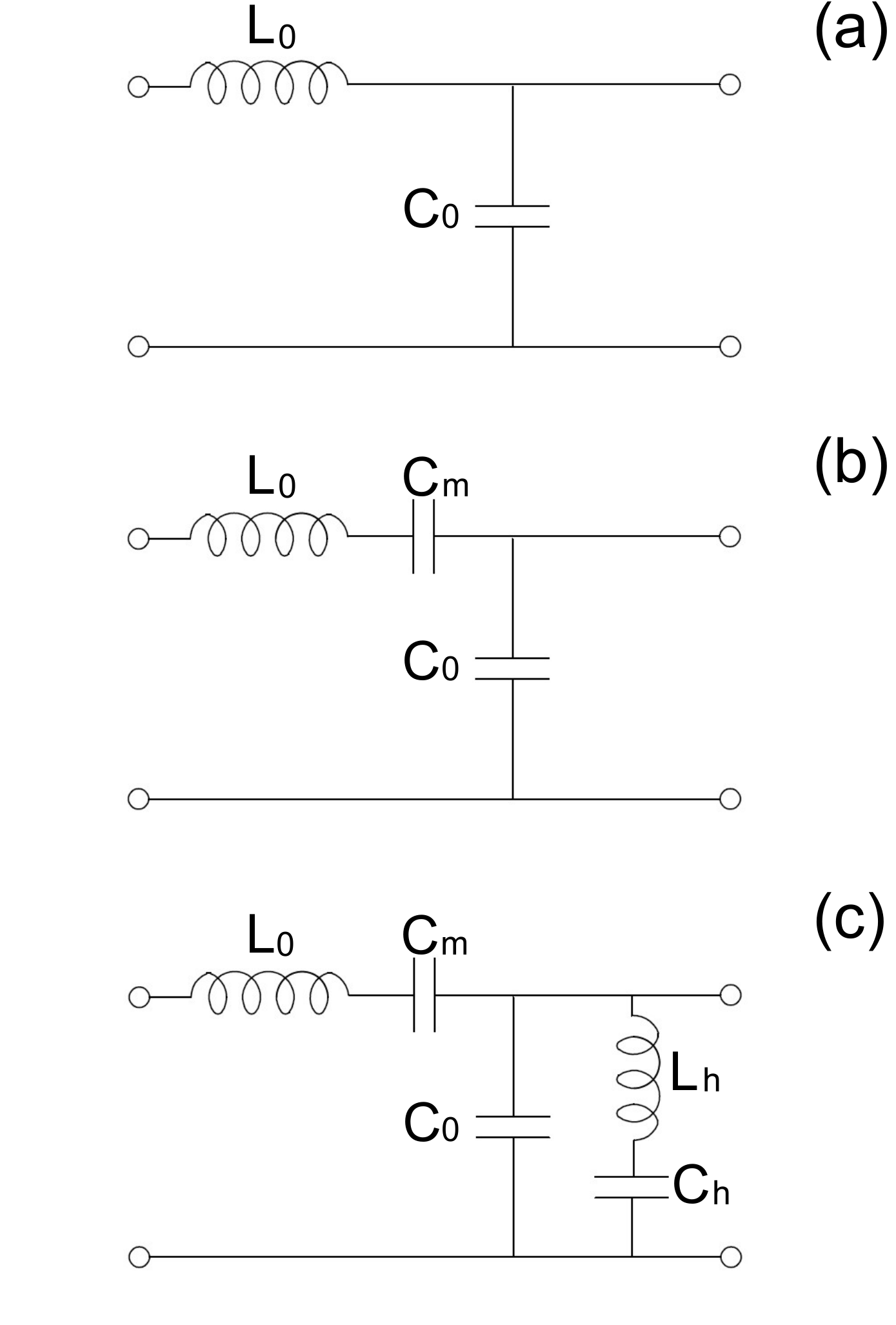}
\end{center}
\caption{(a) Unit of the transmission line that represents a
conventional acoustic medium. (b) Electrical equivalence of a unit
cell of the $\rho$ -NG structure based on an array of membranes. (c)
Representation of the present acoustic metamaterial: the circuit
consists of series connected $L_0$ and $C_m$, and shunt connected
$C_0$, $L_h$, and $C_h$.} \label{fig:circuit}
\end{figure}

Acoustic metamaterials can be described by transmission line models,
as recently demonstrated by Bongard \emph{et al.}~\cite{16}. In
these models, the voltage ($V$), current ($I$), inductor ($L$), and
capacitor ($C$) of the circuits correspond to the pressure ($p$),
volume flow ($q$), mass of air ($M$), and compressibility of the
acoustic media ($1/B$), respectively. In other words, $V
\leftrightarrow p$, $I \leftrightarrow q$, $L \leftrightarrow
M/S^2$, and $C \leftrightarrow \Omega/B$, where $S$ and $\Omega$ are
the cross-section and unit cell volume of the tube respectively. A
plain tube filled with air is represented by a circuit consisting of
an array of series inductors and shunt capacitors, as shown in Fig.
2(a). The constitutive parameters for this medium are the
per-unit-length inductance, $l_0=L_0/d$, and the per-unit-length
capacitance, $c_0=C_0/d$, where $d$ is the length of a unit cell.
These parameters correspond to the mass-density, $\rho_0$, and
compressibility, $B^{-1}_0$, of air, respectively (i.e., $l_0
\leftrightarrow \rho_0$ and $c_0 \leftrightarrow B^{-1}_0$). When an
array of membranes is additionally installed in the tube (making the
resulting structure $\rho$ -NG), the unit of the corresponding
transmission line becomes like that shown is Fig. 2(b)~\cite{16},
where an additional series capacitor $C_m$ represents the membrane.

\begin{figure}
\begin{center}
\includegraphics*[width=1.0\columnwidth]{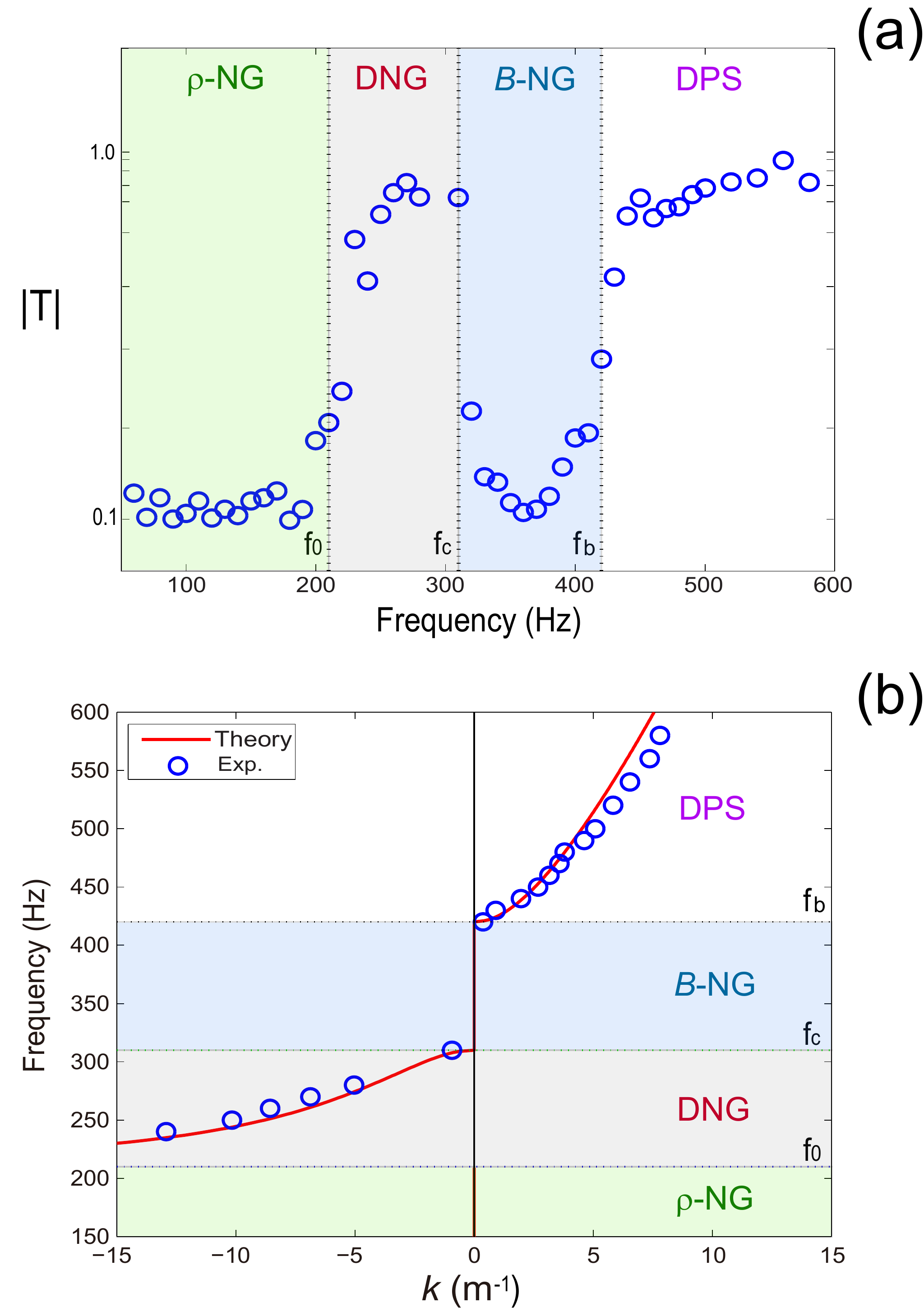}
\end{center}
\caption{(a) Transmission data shows two pass bands and two stop
bands. (b) Wave-vectors are negative in the frequency band
$\omega_0<\omega<\omega_c$ ($f=\omega/2\pi$), and positive above
$\omega_b$.} \label{fig:veolcity}
\end{figure}

The present acoustic composite structure consisting of membranes and
Helmholtz resonators can be analyzed using the circuit shown in Fig.
2(c). The electrical resonator connected in parallel with the
capacitor $C_0$ represents the Helmholtz resonator; the inductor
$L_h$ and capacitor $C_h$ correspond to the mass of the air column
in the neck and the spring action of the cavity volume,
respectively. The generalized telegrapher's equations for the
voltage, $V=V(z)exp(-i\omega t)$ and current waves,
$I=I(z)exp(-i\omega t)$ are expressed as

\begin{equation*} \label{eq:a}
V(z)=-Z'I(z)
\end{equation*}
\begin{equation*} \label{eq:b}
I(z)=-Y'V(z),
\end{equation*}
where $Z'$ and $Y'$ are the transmission line equivalent overall
per-unit-length of impedance and admittance, respectively. For the
circuit in Fig. 2(c), it can be shown that

\begin{equation} \label{eq:c}
Z'=i\omega l_0 \left(1-\frac{\omega_c^2}{\omega^2}\right),
\end{equation}
\begin{equation} \label{eq:d}
Y'=i\omega c_0
\left(1-\frac{F\omega_0^2}{\omega^2-\omega_0^2}\right),
\end{equation}
where $\omega_c=1/\sqrt{L_0C_m}$, $\omega_0=1/\sqrt{L_hC_h}$, and
$F=C_h/C_0$.

The physics related to both the transmission line and the acoustic
metamaterial are identical. There are also one to one
correspondences for all quantities. The constitutive parameters
$l_{eff}$ and $c_{eff}$ for the transmission line, which correspond
to the effective density and modulus of the acoustic system,
respectively, are obtained from the relations $Y'= i\omega c_{eff}$
and $Z'= i\omega l_{eff}$. Explicit expressions are obtained from
Eqs. (3) and (4),

\begin{equation} \label{eq:5}
c_{eff}=c_0\left(1-\frac{F\omega_0^2}{\omega^2-\omega_0^2}\right),
\end{equation}
\begin{equation} \label{eq:6}
l_{eff}=l_0\left(1-\frac{\omega_c^2}{\omega^2}\right).
\end{equation}
It should be noted that Eq. (1) for the $\rho$ -NG medium and (2)
for the B-NG medium, exactly match Eqs. (5) and (6), respectively.
Since Eqs. (5) and (6) describe the transmission line, Eqs. (1) and
(2) are simultaneously applicable to its counterpart, the present
acoustic system.

Using the relations, $k^2(\omega)=\omega^2l_{eff}c_{eff}$ and
$k^2(\omega)=\omega^2\rho_{eff}B_{eff}^{-1}$, the wave-vectors for
the transmission line and the acoustic metamaterial are found to
have the same expression,

\begin{equation} \label{eq:8}
k^2(\omega)=\frac{1}{v_0^2}\frac{(\omega^2-\omega_c^2)(\omega^2-\omega_b^2)}{\omega^2-\omega_0^2},
\end{equation}
where $\omega_b=\omega_0\sqrt{1+F}$, and $v_0$ is the free space
phase velocity; $v_0=1/\sqrt{l_0c_0}$ for the transmission line and
$v_0=1/\sqrt{\rho'B_{0}^{-1}}$ for the acoustic system. Eq. (7) is
identical in form to the dispersion relation for the electromagnetic
metamaterial in Ref. 4.

We observed the transmission and dispersion relation using the
following method. A sound source was installed on one side of the 2
m long metamaterial and an absorber was placed on the other side, so
that the sound energy from the source propagating along the
metamaterial tube was absorbed at the other end. The sound source
was a speaker (RP-HV102, Panasonic) driven by an arbitrary function
generator (33220A, Agilent). The absorber, constructed via a method
similar to that described in Ref. 14 and 15, absorbed most of the
incoming energy and allowed only a small reflection. Thus, the
acoustic wave propagating in the metamaterial behaved as if it
extended to infinity. Transmission data were obtained with by
miniature condenser type microphones (MS-9600, Neosonic) placed in
front of the source and at a position 1.5 m away from the source.
Phase velocity data were obtained by comparing the phases of the
sinusoidal signals from detectors placed at two positions and
separated by a given distance.

The transmission data are shown in Fig. 3(a). There are two stop
bands in the frequency ranges $\omega<\omega_0$ and
$\omega_c<\omega<\omega_b$ where the wave-vector in Eq. (7) is
imaginary. There are also two pass bands, in the frequency ranges
$\omega_0<\omega<\omega_c$ and $\omega_b<\omega$. The phase velocity
$v_{ph}$ was measured as a function of the frequency, and using the
relation $k=\omega/v_{ph}$, the wave-vector data were obtained as
shown in Fig 3(b). The sign of the phase velocity is assigned to be
negative when its direction is opposite that of energy flow. Waves
moving towards the source was observed in the frequency range
$\omega_0<\omega<\omega_c$ (negative phase velocity). In the upper
pass band, $\omega_b<\omega$, the waves moved away from the source
(positive phase velocity). The experimental data are in excellent
agreement with the theoretical expectations, which therefore
confirms the validity of the present analysis using the transmission
line model, because any error in the theory would have caused a
deviation from the observations.

\begin{figure}
\begin{center}
\includegraphics*[width=1.0\columnwidth]{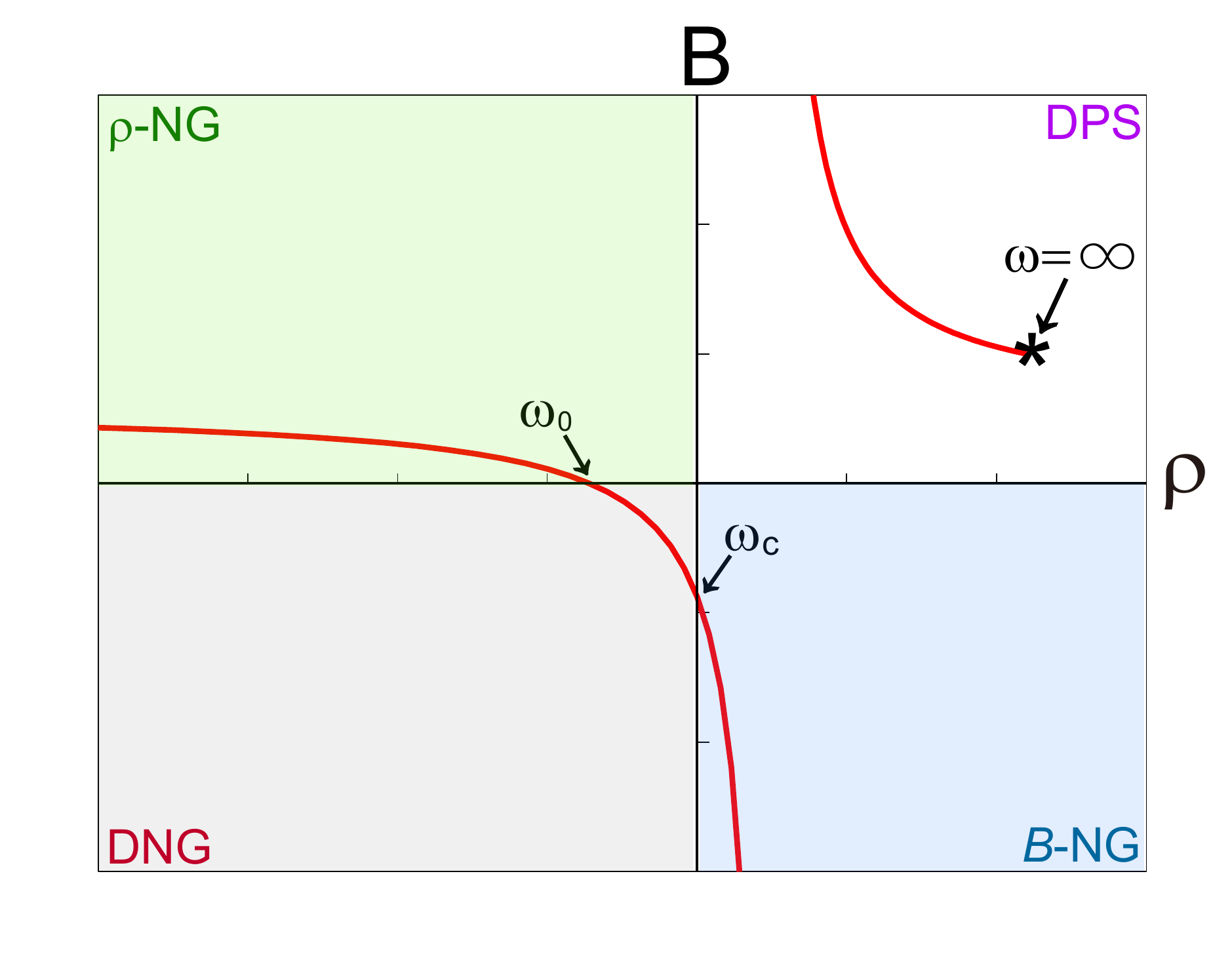}
\end{center}
\caption{Graphical representation of the wave characteristics. At
low frequencies, the curve starts out from the 2nd quadrant and
moves to the right. As the frequency is increased the curve sweeps
through all the four quadrants of the $\rho$ -$B$ plane.}
\label{fig:veolcity1}
\end{figure}

Veselago used $\epsilon$-$\mu$ diagram~\cite{1} to introduce the
concept of single negative and double negative electromagnetic
media, but no investigation has been reported so far on representing
the states of the metamaterial as the points in the $\epsilon$-$\mu$
plane. In all metamaterials, the values of the constitutive
parameters, which determine all the wave characteristics, vary with
the frequency. Therefore we shall call a set of particular values of
the constitutive parameter as an acoustic or optical state of the
medium and represented it as a point in the $\rho$-$B$ or
$\epsilon$-$\mu$ diagram. As the frequency varies, the point moves
in the plane, and the trajectory forms a graph of the available
states. Many of the characteristics of the metamaterial can
conveniently be obtained from the graph.

In our structure, the transition frequencies were observed to be,
$\omega_0=210$ Hz, $\omega_c=310$ Hz, and $\omega_b=420$ Hz.
Substituting these values into the Eqs. (1) and (2), we obtained the
graph in the $\rho$-$B$ plane as shown in Fig. 4. At low
frequencies, the curve starts out from the 2nd quadrant and moves to
the right as the frequency increases until it hits the $\rho$ -axis
at $\omega_0$. For the next frequency band,
$\omega_0<\omega<\omega_c$, the curve passes through the 3rd
quadrant, where, the ratio $\rho/B$ varies from zero to infinity,
indicating a wide range of refractive index. Indeed, we observed the
index $n$ relative to air varying continuously with the frequency
from -0.06 at 310 Hz to -3.7 at 230 Hz (see the wave-vector data in
Fig 3(b)). The line in the fourth quadrant corresponds to the
frequency range, $\omega_c<\omega<\omega_b$. At $\omega_b$, the
modulus diverges to $-\infty$. The states above the frequency
$\omega_b$ are represented by the curve in the 1st quadrant, which
started from $B=\infty$ and approached to the asymptotic value
indicated by the "$\ast$" in the figure. Consequently the curve
swept through all the four quadrants of the $\rho$ -$B$ plane.
Therefore, our structure exhibits $\rho$ -NG, DNG, $B$-NG, and DPS
states in sequence with the frequency.

In summary, we described the fabrication of a double negative
acoustic metamaterial consisting of an interspaced array of
membranes and Helmholtz resonators. An equivalent transmission line
circuit and graphical representations in the $\rho$ -$B$ planes were
used to analyze the structure. Sharp transitions at three
frequencies, $\omega_0$, $\omega_c$, and $\omega_b$, were observed
to change the state of the proposed acoustic metamaterial from
$\rho$ -NG to DNG, $B$-NG, and DPS in sequence. In the DNG band, a
broad refractive index range was observed, which may allow for new
acoustic wave transformation schemes.

This research was supported by Basic Science Research Program
through the National Research Foundation of Korea(NRF) funded by the
Ministry of Education, Science and Technology (NRF 2010-0012562).

\end{document}